\documentclass[aps,prc,superscriptaddress,floats,floatfix,reprint,reprintnumbers]{revtex4}
\usepackage{blindtext}
\pdfoutput=1
\usepackage{epsfig}
\usepackage{graphicx}
\usepackage{dcolumn}
\usepackage{bm}
\usepackage{booktabs}
\usepackage{color}
\usepackage{mathrsfs}
\usepackage{ulem}
\usepackage{epstopdf}
\graphicspath{{plots/}}
\usepackage{verbatim}
\usepackage{indentfirst}
\usepackage{float}
\usepackage[colorlinks,
           bookmarks=true,
           linkcolor=blue,
           urlcolor=blue,
            anchorcolor=black,
            citecolor=blue]{hyperref}
\usepackage[bf,raggedright,indentafter]{titlesec}
\usepackage{rotating}
\usepackage{booktabs}
\usepackage{tabularx}
\usepackage{makecell}
\usepackage{hypcap}
\usepackage{subfigure}%
\usepackage{tabularx}

\usepackage{array}

\usepackage{amsmath}

\begin{document}

\title{\Large Four-body Semileptonic Decays  $B\to D^*P\ell^+\nu_\ell$ with the SU(3) Flavor Symmetry}
\author{Meng-Yuan Wan,~~~Yuan-Guo Xu$^{\dag}$,~~~Qi-Lin Jia,~~~Yue-Xin Liu,~~~Yi-Jie Zhang\\
 {\scriptsize College of Physics and Communication Electronics, JiangXi Normal University, NanChang, JiangXi 330022, China}\\
 $^\dag${\scriptsize Corresponding author. E-mail addresses:yuanguoxu@jxnu.edu.cn}
}

\begin{abstract}\vspace{1cm}
We present a complete study of  the $B\to D^*P\ell^+\nu_\ell~(\ell=e,\mu,\tau)$  decays with  the non-resonant, the charmed axial vector resonant and  the charmed  tensor resonant contributions
by using the SU(3) flavor symmetry.
Relevant amplitude relations between different decay modes are obtained by the SU(3) flavor symmetry. We then predict non-measured  branching ratios of the $B\to D^*P\ell^+\nu_\ell$  decays with
the non-resonant and the charmed  resonant contributions  by using present experimental data of the $B\to D^*P\ell'^+\nu_{\ell'}~(\ell'=e,\mu)$  decays within $2\sigma$ errors.
We have found that  $B^{0,+}\to D^*\eta\ell^+\nu_\ell$, $B^{0,+}\to D^*\eta'\ell^+\nu_\ell$, $B^{0}_s\to D^*_s \eta\ell^+\nu_\ell$, $B^{0}_s\to D^*_s\eta'\ell^+\nu_\ell$
and $B^{0,+}\to D^{*}_sK\ell^+\nu_\ell$ decays only receive non-resonant contributions. Decays $B^0_s\to D_s^{*-}\pi^0\ell^+\nu_\ell$ only receive the $D'_{s1}$ resonant contributions.
Other decays receive all three kinds of contributions, and three kinds of  contributions are important in most of decays.

\end{abstract}
\maketitle

\section{Introduction}
Decays $B\to D^*\pi\ell'^+\nu_{\ell'}$, $B^+\to D^{*-}_sK^+\ell'^+\nu_{\ell'}$ and their resonant decays
 $B\to D^{(')}_1(\to D^*\pi)\ell'^+\nu_{\ell'}$, $B\to D^*_2(\to D^*\pi)\ell'^+\nu_{\ell'}$   have been measured by BABAR, Belle and CLEO, for examples, Refs. \cite{ParticleDataGroup:2022pth,Belle:2022yzd,BaBar:2010ner}.
Present measurements of the $B\to D^*P\ell'^+\nu_{\ell'}$ decays give us an opportunity to additionally test theoretical approaches and to predict many  non-measured $B\to D^*P\ell^+\nu_\ell$ decays.
And the predictions of the four-body semileptonic  decays $B\to D^*P\ell^+\nu_\ell$ can be tested  by LHC and Belle-II in the near future.

For the semileptonic decays, since leptons do not participate in the strong interaction, the weak and strong dynamics
can be separated, the theoretical description of the semileptonic  decays are relatively simple. Moreover,  the  experimental backgrounds of  the semileptonic decays are cleaner than ones of the hadronic decays.
The hadronic transition form factors, which are crucial for testing the theoretical calculations of the involved strong interaction, contain all the strong dynamics in the initial and final hadrons.
The analytic structure of the $B\to D^*P$ form factors is more complicated than for the $B\to D^*$ or $B\to P$ form factors, as similarly analysed in Ref. \cite{Feldmann:2018kqr}, and the resonance states also involve
in the $B\to D^*P\ell^+\nu_\ell$ decays.
There are several methods to calculate the form factors, such as the chiral perturbation theory \cite{Kang:2013jaa}, the unitarized chiral perturbation theory \cite{Shi:2017pgh,Shi:2020rkz},
the light-cone sum rules \cite{Sekihara:2015iha,Cheng:2017smj,Hambrock:2015aor},  and the QCD factorization \cite{Boer:2016iez}.  Nevertheless, due to our poor understanding
of hadronic interactions, the evaluations of the form factors  are difficult and  often plugged with large uncertainties.

In the absence of reliable calculations, the symmetry analysis can provide very useful information about the decays.
SU(3) flavor symmetry, which is independent of the detailed dynamics, provides us an opportunity to relate different decay modes,
nevertheless, it cannot determine the sizes of the amplitudes or the form factors by itself.
However, if experimental data are enough, one may use the data to extract the form factors.
Even though it is only an approximate symmetry due to up, down, and strange quarks having different masses, it still gives some valuable information about the decays.
SU(3) flavor symmetry has been widely used to study hadron decays, including $b$-hadron decays  \cite{He:1998rq,He:2000ys,Fu:2003fy,Hsiao:2015iiu,He:2015fwa,He:2015fsa,Deshpande:1994ii,Gronau:1994rj,Gronau:1995hm,Shivashankara:2015cta,Zhou:2016jkv,Cheng:2014rfa,Wang:2021uzi,Wang:2020wxn},   $c$-hadron decays  \cite{Wang:2021uzi,Wang:2020wxn,Grossman:2012ry,Pirtskhalava:2011va,Savage:1989qr,Savage:1991wu,Altarelli:1975ye,Lu:2016ogy,Geng:2017esc,Geng:2018plk,Geng:2017mxn,Geng:2019bfz,Wang:2017azm,Wang:2019dls,Wang:2017gxe,Muller:2015lua}, and light hadron decays \cite{Wang:2019alu,Wang:2021uzi,Xu:2020jfr,Chang:2014iba,Zenczykowski:2005cs,Zenczykowski:2006se,Cabibbo:1963yz}.

SU(3) flavor  breaking effects  play a key role in the precise theoretical predictions of the observables and a precise test of the the
unitarity of the CKM matrix.
SU(3) flavor  breaking effects  due to the mass difference  of $u,d,s$ quarks have also been studied, for instance,  in Refs. \cite{Martinelli:2022xir,Wang:2022fbk,Imbeault:2011jz,Wu:2005hi,Dery:2020lbc,Sasaki:2008ha,Pham:2012db,Geng:2018bow,Flores-Mendieta:1998tfv,Cheng:2012xb,Xu:2013dta,He:2014xha,Yang:2015era}.   The  SU(3) flavor  breaking effects  have been found to be small, for examples, in light hyperon
semileptonic decays \cite{Pham:2012db,Yang:2015era} and $B\to D\ell\nu_\ell$  decays \cite{Martinelli:2022xir}.  Nevertheless, they might be given larger effects in some other processes, for examples, $D^0\to \pi^0e^-\nu_e$ decay \cite{Wang:2022fbk} and   $B\to D^*\ell\nu_\ell$  decays \cite{Martinelli:2022xir}.  So   the  SU(3) flavor  breaking effects should be considered for  precise theoretical predictions.
Just, if considering the SU(3) flavor  breaking effects, extra non-perturbative parameters appear, and
one needs more experimental data to determine  relevant  non-perturbative parameters.  At present, there are not many relevant experimental data of  the $B\to D^*P\ell^+\nu_\ell$ decays. As more and more accurate data are collected in the future, one may study the  SU(3) flavor  breaking effects in the the $B\to D_1/D'_1/D^*_2 \ell^+\nu_\ell$ and $B\to D^*P\ell^+\nu_\ell$ decays.
So only the SU(3) flavor  symmetry contributions will be  analyzed  in this work.

Some four-body semileptonic decays $B/D\to P_1 P_2 \ell^+\nu_\ell$ and $B\to D^{(*)} P \ell^+\nu_\ell$ have been studied, for instance, in Refs. \cite{Cheng:1993ah,Tsai:2021ota,Feldmann:2018kqr,Kim:1999gm,Kim:1998nn,LeYaouanc:2018zyo,Boer:2016iez,Hambrock:2015aor,Faller:2013dwa,Shi:2021bvy,Achasov:2020qfx,Wiss:2007mr,Wang:2016wpc,Achasov:2021dvt,Gustafson:2023lrz}.
In this work,  we will  study the $B\to D^*P\ell^+\nu_\ell$ decays  with  the non-resonant, the charmed axial vector resonant and  the charmed  tensor resonant contributions by the SU(3) flavor symmetry.
Firstly, we will construct the hadronic amplitude or the form factor  relations between different decay modes and use the available data to extract them. Then, we will predict the not-yet-measured modes for further tests in experiments and analyze the contributions with the non-resonance.

This paper is organized as follows. In Sec. II, we  calculate the nonresonant contributions of the $B\to D^* P\ell\nu_\ell$ decays. In Sec. III, we   present
the charmed axial vector resonant and  the charmed  tensor resonant contributions of  $B \rightarrow  D^* P\ell^+\nu_\ell$ decays.  Finally, we summarize our work in Sec. IV.


\section{Non-resonant $B\to D^* P\ell\nu_\ell$ decays}

\subsection{Decay amplitudes}

The four-body semileptonic  decays $B \rightarrow D^*P \ell^+\nu_\ell$ are generated by   $\bar{b}\rightarrow \bar{c} \ell^+\nu_\ell$ transition, and the effective Hamiltonian   is
\begin{eqnarray}
\mathcal{H}_{eff}(\bar{b}\rightarrow \bar{c} \ell^+\nu_\ell)=\frac{G_F}{\sqrt{2}}V_{cb}\bar{b}\gamma^\mu(1-\gamma_5)\bar{c}~\bar{\nu_\ell}\gamma_\mu(1-\gamma_5)\ell,\label{Heff}
\end{eqnarray}
where $G_F$ is the Fermi constant, $V_{cb}$ is the CKM matrix element. Decay amplitudes of the $B \rightarrow D^*P \ell^+\nu_\ell$ decays can be written as
\begin{eqnarray}
\begin{aligned}
\mathcal{A}(B \rightarrow D^*P  \ell^+\nu_\ell)&=\langle D^*(k_1)P(k_2) \ell^+(q_1)\nu_\ell(q_2)| \mathcal{H}_{eff}(\bar{b}\rightarrow \bar{c} \ell^+\nu_\ell) |B(p_B)\rangle\\
&=\frac{G_F}{\sqrt{2}}V_{cb} L_{\mu}H^{\mu},\label{Am}
\end{aligned}
\end{eqnarray}
where  $L_\mu=\bar{\nu_\ell}\gamma_{\mu}(1-\gamma_5)\ell$ is the leptonic  charged current, and $H^\mu=\langle D^*(k_1)P(k_2)|\bar{c}\gamma^\mu(1-\gamma_5)\bar{b}|B(p_B)\rangle$ is the hadronic matrix element.

Usually, the hadronic matrix element $H^\mu$ can be obtained in terms of the form factors of the $B \rightarrow D^*P$. Similar to  $B\to PP$  form factors  given in Ref. \cite{Boer:2016iez},
the $B \rightarrow D^*P$ form factors are defined by
\begin{eqnarray}
\langle D^*(k_1)P(k_2)|\bar{c}\gamma^\mu \bar{b}|B(p_B)\rangle&=&iF_\perp \frac{1}{\sqrt{k^2}}q^\mu_{\perp},  \\
-\langle D^*(k_1)P(k_2)|\bar{c}\gamma^\mu \bar{b}|B(p_B)\rangle&=&F_t \frac{q^\mu}{\sqrt{q^2}}+F_0 \frac{2\sqrt{q^2}}{\sqrt{\lambda}}k^\mu_{0}+F_\parallel \frac{1}{\sqrt{k^2}}\bar{k}^\mu_{\parallel}, \label{Eq:DFFB2PP}
\end{eqnarray}
where $k\equiv k_1+k_2$, $q\equiv q_1+q_2$, $\lambda=\lambda(m_{D^*}^2,q^2,k^2)$ with $\lambda(a,b,c)= a^2+b^2+c^2-2ab-2bc-2ac$, and $F_0,F_t,F_\perp,F_\parallel$ are the form factors of $B \rightarrow D^*P$. In addition,  $q^\mu_{\perp}$, $k^\mu_{0}$ and $\bar{k}^\mu_{\parallel}$   are defined in Ref. \cite{Boer:2016iez}.

The differential branching ratios of the nonresonant $B \rightarrow D^*P \ell^+\nu_\ell$ decays can be written as
\begin{eqnarray}
    \frac{d\mathcal{B}(B \rightarrow D^*P  \ell^+\nu_\ell)_N}{dq^2 dk^2}=\frac{1}{2}\tau_{B}|\mathcal{N}|^2\beta_\ell(3-\beta_\ell)|H_N|^2,
\end{eqnarray}
with
\begin{eqnarray}
{|\mathcal{N}|}^2&=&G^2_F {|V_{cb}|}^2 \frac{\beta_\ell q^2 \sqrt{\lambda}}{3 \times 2^{10} \pi^5 m^3_B}\,,\quad \text{with}\quad
\beta_\ell=1-\frac{m^2_\ell}{q^2}\,. \nonumber \\
|H_N|^2&=& |F_0|^2+\frac{2}{3}(|F_{\parallel}|^2+|F_\perp|^2)+\frac{3m_\ell^2}{q^2(3-\beta_\ell)}|F_t|^2,\label{Eq:DB2PPlvdbr}
\end{eqnarray}
where $\tau_{M}$($m_{M}$) is lifetime(mass) of $M$ particle. The ranges of integration are given by  $(m_{D^*}+m_{P})^2\leq k^2\leq(m_{B}-m_\ell)^2$ and $m_\ell^2\leq q^2 \leq(m_{B}-\sqrt{k^2})^2$.

The calculations of the $F_0,F_t,F_\perp,F_\parallel$ form factors are difficult. If we ignore $|F_t|^2$ term since it is proportional to $m_\ell^2$ and it is small when $\ell=e,\mu$, $|H_N|^2$ is only include the hadronic part.
Noted that although $|F_t|^2$ term might be large when $\ell=\tau$, it is difficult to estimate its contribution in this work, so we still ignore it.
Then $|H_N|^2$, which only includes hadronic part, can be  related by the SU(3) flavor symmetry/breaking of $u$, $d$, $s$ quarks.

\subsection{Hadronic amplitudes based on the SU(3) flavor analysis}
Since the SU(3) flavor analysis is based on the SU(3) flavor group, we will give relevant meson multiplets first. Bottom pseudoscalar  triplets $B_i$,  charm vector  triplets $D^*_i$,  and light pseudoscalar  octets and singlets $P^i_j$ under the SU(3) flavor symmetry of u, d, s quarks are
\begin{eqnarray}
B_i &=& \Big(B^+(\bar{b} u),B^0(\bar{b} d),B^0(\bar{b} s)\Big),\\
D^*_i&=&\Big( \overline{D}^{*0}(\bar{c}u),~D^{*-}(\bar{c}d),~D^{*-}(\bar{c}s)\Big),\\
 P^i_j&=&\left(\begin{array}{cccc}
\frac{\pi^0}{\sqrt{2}}+\frac{\eta_8}{\sqrt{6}}+\frac{\eta_1}{\sqrt{3}}& \pi^+ & K^+\\
\pi^- &-\frac{\pi^0}{\sqrt{2}}+\frac{\eta_8}{\sqrt{6}}+\frac{\eta_1}{\sqrt{3}} & K^0 \\
K^- & \overline{K}^0 &-\frac{2\eta_8}{\sqrt{6}}+\frac{\eta_1}{\sqrt{3}}\\ \end{array}\right)\,,
\end{eqnarray}
where $i,j=1,2,3$ for $u$, $d$, $s$ quarks, the $\eta$ and $\eta'$  are mixtures of $\eta_1=\frac{u\bar{u}+d\bar{d}+s\bar{s}}{\sqrt{3}}$ and $\eta_8=\frac{u\bar{u}+d\bar{d}-2s\bar{s}}{\sqrt{6}}$ with the mixing angle $\theta_P$
\begin{eqnarray}
\left(\begin{array}{c}
\eta\\
\eta'
\end{array}\right)\,
=
\left(\begin{array}{cc}
\mbox{cos}\theta_P&-\mbox{sin}\theta_P\\
\mbox{sin}\theta_P&\mbox{cos}\theta_P
\end{array}\right)\,\left(\begin{array}{c}
\eta_8\\
\eta_1
\end{array}\right)\,.
\end{eqnarray}\label{Eq:etamix}
And $\theta_P=[-20^\circ,-10^\circ]$ from  the Particle Data Group (PDG) \cite{ParticleDataGroup:2022pth} will be used in our numerical  analysis.

Charm axial vector $A$ and tensor $T$  contribute to these processes as resonances, which will be discussed next section.  There are two types of
P-wave charm axial-vector mesons with different  quantum
numbers $J^{PC}=1^{++}$ and $J^{PC}=1^{+-}$. The charm axial vector triplets are
\begin{eqnarray}
D'_{1i}&=&\Big( \overline{D}'_1(2430)^0,~D'_1(2430)^-,~D'_{s1}(2460)^-\Big)~~~~~~~~~~~\mbox{for}~~~~~~J^{PC}=1^{++},\\
D_{1i}&=&\Big( \overline{D}_1(2420)^0,~D_1(2420)^-,~D_{s1}(2536)^-\Big)~~~~~~~~~~~\mbox{for}~~~~~~J^{PC}=1^{+-}.
\end{eqnarray}
The charm  tensor triplet is
\begin{eqnarray}
D_{2i}&=&\Big( \overline{D}^*_2(2460)^0,~D^*_2(2460)^-,~D^*_{s2}(2573)^-\Big).
\end{eqnarray}

\begin{figure}[t]
\begin{center}
\includegraphics[scale=0.6]{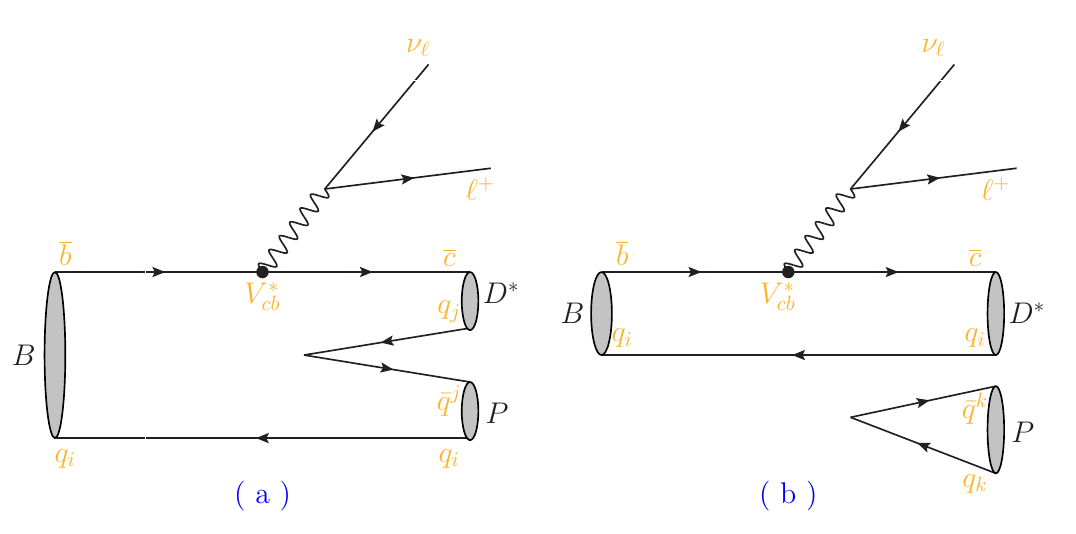}
\end{center}
\caption{Diagrams of the non-resonant  $B\to D^*P\ell^+\nu_\ell$ decays. }\label{Fig1}
\end{figure}

Feynman diagrams for the non-resonant $B\to D^*P\ell^+\nu_\ell$ decays
are displayed in Fig. \ref{Fig1}.
The leptonic charged current  is  invariant under the SU(3) flavor symmetry, and the hadronic matrix element can be parameterized by the SU(3) flavor symmetry.

The decay amplitudes of the non-resonant $B\rightarrow D^*P\ell^+\nu_\ell$ decays  in Eq. (\ref{Am}) can be  transformed as
\begin{eqnarray}
\mathcal{A}(B \rightarrow D^*P  \ell^+\nu_\ell)_N&=&\frac{G_F}{\sqrt{2}}V_{cb}^* H(B \rightarrow D^*P )_N~\bar{\nu}_\ell\gamma_\mu(1-\gamma_5)\ell,\\
H(B\to D^*P)_N &=&a_{1}B^iP_i^j{D^*}_j+a_{2}B^i{D^*}_iP^k_k,\label{Eq:HND2PPlv}
\end{eqnarray}
where $a_{1,2}$ are the nonperturbative coefficients under the SU(3) flavor symmetry.

The hadronic amplitudes of the  non-resonant $B\to D^*P\ell^+\nu_\ell$ decays  are
given in  Tab. \ref{Tab:HN},
 in which we can see the amplitude relations of them. As given in Fig. \ref{Fig1} (b), the $a_2$ term in Eq. (\ref{Eq:HND2PPlv}) is suppressed by the Okubo-Zweig-Iizuka (OZI) rule \cite{Okubo:1963fa,Lipkin:1986bi,Lipkin:1996ny}, and it only appears in the decays with $\eta,\eta'$ final states.
 If ignoring the OZI suppressed $a_2$ term,  all hadronic amplitudes   can be related by only one nonperturbative coefficient $a_1$.
\begin{table}[t]
\renewcommand\arraystretch{1.3}
\tabcolsep 0.2in
\centering
\caption{The hadronic amplitudes  $H(B\to D^*P)_N$ for  $B\to D^*P\ell^+\nu_\ell$ decays based on the SU(3) flavor symmetry.
}\vspace{0.1cm}
{\footnotesize
\begin{tabular}{lc|lc}  \hline\hline
~~~Decay modes                                      & Hadronic  amplitudes                                                             & ~~~Decay modes                                     & Hadronic  amplitudes\\\hline
$B^+\to \overline{D}^{*0}\pi^0\ell^+\nu_\ell$       &$\frac{a_1}{\sqrt{2}}$                                                            &$B^0\to \overline{D^*}^0\pi^-\ell^+\nu_\ell$        &$a_1$                                                                           \\
$B^+\to \overline{D}^{*0}\eta\ell^+\nu_\ell$        &$\frac{a_1}{\sqrt{6}}cos\theta_P-\frac{a_1+3a_2}{\sqrt{3}}sin\theta_P$            & $B^0\to D^{*-}\pi^0\ell^+\nu_\ell$                 &$-\frac{a_1}{\sqrt{2}}$                                                        \\
$B^+\to \overline{D}^{*0}\eta'\ell^+\nu_\ell$       &$\frac{a_1}{\sqrt{6}}sin\theta_P+\frac{a_1+3a_2}{\sqrt{3}}cos\theta_P$            &$B^0\to  D^{*-}\eta \ell^+\nu_\ell$                 &$\frac{a_1}{\sqrt{6}}cos\theta_P-\frac{a_1+3a_2}{\sqrt{3}}sin\theta_P $   \\
$B^+\to D^{*-}\pi^+\ell^+\nu_\ell$                  &$a_1$                                                                             & $B^0\to D^{*-}\eta'\ell^+\nu_\ell$                 &$\frac{a_1}{\sqrt{6}}sin\theta_P+\frac{a_1+3a_2}{\sqrt{3}}cos\theta_P $    \\
$B^+\to D^{*-}_sK^+\ell^+\nu_\ell$                  &$a_1$                                                                             & $B^0\to D^{*-}_sK^0\ell^+\nu_\ell$                 &$a_1$                                                                           \\
$B^0_s\to \overline{D^*}^0K^-\ell^+\nu_\ell$        &$a_1$                                                                             &$B^0_s\to D^{*-}_s\eta \ell^+\nu_\ell$              &$-\frac{2a_1}{\sqrt{6}}cos\theta_P-\frac{a_1+3a_2}{\sqrt{3}}sin\theta_P $     \\
$B^0_s\to D^{*-}\overline{K}^0\ell^+\nu_\ell$          &$a_1$                                                                             &$B^0_s\to D^{*-}_s\eta'\ell^+\nu_\ell$              &$-\frac{2a_1}{\sqrt{6}}sin\theta_P+\frac{a_1+3a_2}{\sqrt{3}}cos\theta_P $     \\
\hline
\end{tabular}\label{Tab:HN}}
\end{table}

\subsection{Numerical results for the non-resonant $B\rightarrow D^*P\ell^+\nu_\ell$ decays}
The theoretical input
parameters  and the
experimental data within the $2\sigma$ errors from PDG \cite{ParticleDataGroup:2022pth}
will be used in our numerical analysis.
The four-body semileptonic $B^+\to D^{*-}\pi^+\ell'^+\nu_{\ell'}$ and $B^0\to D^{*0}\pi^+\ell'^+\nu_{\ell'}$ decays  have been well measured \cite{ParticleDataGroup:2022pth}, and they with the $2\sigma$ errors are
\begin{eqnarray}
\mathcal{B}(B^+\to D^{*-}\pi^+\ell'^+\nu_{\ell'})_{all}&=& (6.0\pm0.8)\times10^{-3},\label{EBrTBu2DVpilv}\\
\mathcal{B}(B^+\to D^{*-}\pi^+\ell'^+\nu_{\ell'})_{\overline{D}_1^0}&=& (3.03\pm0.40)\times10^{-3},\label{EBrDABu2DVpilv}\\
\mathcal{B}(B^+\to D^{*-}\pi^+\ell'^+\nu_{\ell'})_{\overline{D}'^0_1}&=& (2.7\pm1.2)\times10^{-3},\label{EBrDApBu2DVpilv}\\
\mathcal{B}(B^+\to D^{*-}\pi^+\ell'^+\nu_{\ell'})_{\overline{D}^{*0}_2}&=& (1.01\pm0.48)\times10^{-3},\label{EBrDTBu2DVpilv}\\
\mathcal{B}(B^0\to D^{*0}\pi^-\ell'^+\nu_{\ell'})_{all}&=& (5.8\pm1.6)\times10^{-3},\label{EBrTBd2DVpilv}\\
\mathcal{B}(B^0\to D^{*0}\pi^-\ell'^+\nu_{\ell'})_{D_1^-}&=& (2.80\pm0.56)\times10^{-3},\label{EBrDABd2DVpilv}\\
\mathcal{B}(B^0\to D^{*0}\pi^-\ell'^+\nu_{\ell'})_{D'^-_1}&=& (3.1\pm1.8)\times10^{-3},\label{EBrDApBd2DVpilv}\\
\mathcal{B}(B^0\to D^{*0}\pi^-\ell'^+\nu_{\ell'})_{D^{*-}_2}&=& (0.68\pm0.24)\times10^{-3},\label{EBrDTBd2DVpilv}
\end{eqnarray}
where $\mathcal{B}_{all,R}$ denotes the total and $R$ resonant branching ratios. From Eqs. (\ref{EBrTBu2DVpilv}-\ref{EBrDTBd2DVpilv}),
we obtain the upper limits of non-resonant branching ratios within  the $2\sigma$ errors
\begin{eqnarray}
\mathcal{B}(B^+\to D^{*-}\pi^+\ell'^+\nu_{\ell'})_N&\leq& 2.14 \times10^{-3},\label{EBrNBu2DVpilv}\\
\mathcal{B}(B^0\to \overline{D}^{*0}\pi^-\ell'^+\nu_{\ell'})_N&\leq& 3.42 \times10^{-3}.\label{EBrNBd2DVpilv}
\end{eqnarray}

In addition, $B^+\to D^{*-}_sK^+\ell'^+\nu_{\ell'}$ has also been measured \cite{ParticleDataGroup:2022pth}
\begin{eqnarray}
\mathcal{B}(B^+\to D^{*-}_sK^+\ell'^+\nu_{\ell'})_{all}&=& (2.9\pm3.8)\times10^{-4}. \label{EBrTBu2DVsKlv}
\end{eqnarray}
Since there are no phase spaces for $\overline{D}_1(2420)^0/\overline{D}'_1(2430)^0/\overline{D}_2(2460)^0\to D^{*-}_sK^+$ decays, $\mathcal{B}(B^+\to D^{*-}_sK^+\ell'^+\nu_{\ell'})_{all}$ will be
considered as only non-resonant branching ratios $\mathcal{B}(B^+\to D^{*-}_sK^+\ell'^+\nu_{\ell'})_N$.  The error of $\mathcal{B}(B^+\to D^{*-}_sK^+\ell'^+\nu_{\ell'})_{all}$ is very large,
all range, that  is less than $6.7\times10^{-4}$, are allowed within  $2\sigma$ error bars.
$\mathcal{B}(B^+\to D^{*-}_sK^+\ell'^+\nu_{\ell'})_N$ as well as   the upper limits of $\mathcal{B}(B^+\to D^{*-}\pi^+\ell'^+\nu_{\ell'})_N$ and $\mathcal{B}(B^0\to \overline{D}^{*0}\pi^-\ell'^+\nu_{\ell'})_N$
in  Eqs. (\ref{EBrNBu2DVpilv}-\ref{EBrTBu2DVsKlv}) will be used to constrain the nonperturbative coefficient  $a_1$.  Present non-resonant data only can give the upper limit $|a_1|\leq20.70$.
And then one can obtain upper limit predictions of other  non-resonant branching ratios, which have not been measured or well measured.

\begin{table}[htb]
\renewcommand\arraystretch{1.25}
\tabcolsep 0.2in
\centering
\caption{Branching ratios  for  $B\to D^*P\ell^+\nu_\ell$ decays due to $ \bar{b}\to \bar{c}\ell^+\nu_\ell$ within  $2\sigma$ errors. The unit is $10^{-4}$ for all branching ratios. $\mathcal{B}_{[R]}$ denotes the R resonant branching ratios. $^e$denotes experimental data within $2\sigma$ errors.  }\vspace{0.1cm}
{\small
\begin{tabular}{lcccc}  \hline\hline
~~~~~~~~~~                                                   &~~$\mathcal{B}_{N}$  &~~~~~~~$\mathcal{B}_{D^{(')}_1}$                                 &~~~~~~ $\mathcal{B}_{D_2^*}$                    &$\mathcal{B}_{all}$  \\\hline
$\mathcal{B}(B^+_u\to D^{*-}\pi^+\ell'^+\nu_{\ell'})$                       &$\leq14.17$         &$^{34.38\pm4.62_{[D'^{0}_1]}}_{30.30\pm4.00_{[D^{0}_1]}}$        &$8.17\pm1.52_{[D^{*0}_2]}$          &$65.54\pm2.46 $               \\
                                                                            &$\leq21.4^e$         & $^{27.00\pm12.00^e_{[D'^{0}_1]}}_{30.30\pm4.00^e_{[D^{0}_1]}}$ &$10.10\pm4.80^e_{[D^{*0}_2]}$       &$60.00\pm8.00^e$         \\\hline
$\mathcal{B}(B^+_u\to D^{*-}_sK^+\ell'^+\nu_{\ell'})$                       &$^{\leq6.70}_{(2.90\pm3.80)^e}$       &$\cdots $                                      &$\cdots $                           &$\cdots$               \\\hline
$\mathcal{B}(B^+_u\to \overline{D}^{*0}\pi^0\ell'^+\nu_{\ell'})$            &$\leq7.17$         &$^{17.44\pm2.34_{[D'^{0}_1]}}_{15.35\pm2.03_{[D^{0}_1]}}$         &$4.31\pm0.80_{[D^{*0}_2]}$          &$40.59\pm8.18 $               \\\hline
$\mathcal{B}(B^+_u\to \overline{D}^{*0}\eta\ell'^+\nu_{\ell'})$             &$\leq2.47$         &$\cdots $                                                         &$\cdots $                           &$\cdots $               \\\hline
$\mathcal{B}(B^+_u\to \overline{D}^{*0}\eta'\ell'^+\nu_{\ell'})$            &$\leq0.80$         &$\cdots $                                                         &$ \cdots$                           &$\cdots $               \\\hline
$\mathcal{B}(B^0_d\to \overline{D}^{*0}\pi^-\ell'^+\nu_{\ell'})$            &$\leq13.32$         &$^{32.37\pm4.33_{[D'^{-}_1]}}_{28.40\pm3.97_{[D^{-}_1]}}$        &$7.82\pm1.38_{[D^{*-}_2]}$          &$67.07\pm6.93 $               \\
                                                                            &$\leq34.20^e$       &$^{31.00\pm18.00^e_{[D'^{-}_1]}}_{28.00\pm5.60^e_{[D^{-}_1]}}$   &$6.80\pm2.40^e_{[D^{*-}_2]}$        &$58.00\pm16.00^e $               \\\hline
$\mathcal{B}(B^0_d\to D^{*-}_sK^0\ell'^+\nu_{\ell'})$                       &$\leq6.21$         &$\cdots $                                                         &$\cdots $                           &$\cdots $               \\\hline
$\mathcal{B}(B^0_d\to D^{*-}\pi^0\ell'^+\nu_{\ell'})$                       &$\leq6.67$         &$^{16.11\pm2.15_{[D'^{-}_1]}}_{14.14\pm1.98_{[D^{-}_1]}}$         &$3.83\pm0.68_{[D^{*-}_2]}$          &$37.17\pm7.50 $               \\\hline
$\mathcal{B}(B^0_d\to D^{*-}\eta\ell'^+\nu_{\ell'})$                        &$\leq2.27$         &$\cdots $                                                         &$\cdots $                           &$\cdots $               \\\hline
$\mathcal{B}(B^0_d\to D^{*-}\eta'\ell'^+\nu_{\ell'})$                       &$\leq0.74$         &$\cdots $                                                         &$\cdots $                           &$\cdots $               \\\hline
$\mathcal{B}(B^0_s\to \overline{D}^{*0}K^-\ell'^+\nu_{\ell'})$              &$\leq8.82$         &$35.60\pm15.40_{[D^{-}_{s1}]}$                                    &$1.71\pm0.51_{[D^{*-}_{s2}]}$       &$41.70\pm20.03 $               \\\hline
$\mathcal{B}(B^0_s\to D^{*-}\overline{K}^0\ell'^+\nu_{\ell'})$              &$\leq8.71$         &$29.67\pm13.61_{[D^{-}_{s1}]}$                                    &$1.30\pm0.40_{[D^{*-}_{s2}]}$       &$35.40\pm17.90$               \\\hline
$\mathcal{B}(B^0_s\to D_s^{*-}\pi^0\ell'^+\nu_{\ell'})$                     &$\cdots $          &$25.45\pm14.11_{[D'^{-}_{s1}]}$                                   &$\cdots$                            &$\cdots $                \\\hline
$\mathcal{B}(B^0_s\to D_s^{*-}\eta\ell'^+\nu_{\ell'})$                      &$\leq3.25$         &$\cdots$                                                          &$\cdots$                            &$\cdots $                \\\hline
$\mathcal{B}(B^0_s\to D_s^{*-}\eta'\ell'^+\nu_{\ell'})$                     &$\leq1.94$         &$\cdots$                                                          &$\cdots$                            &$\cdots$           \\\hline\hline
$\mathcal{B}(B^+_u\to D^{*-}\pi^+\tau^+\nu_\tau)$                       &$\leq1.68$          &$^{5.85\pm0.89_{[D'^{0}_1]}}_{5.12\pm0.76_{[D^{0}_1]}}$         &$5.35\pm1.53_{[D^{*0}_2]}$          &$17.07\pm3.64 $               \\\hline
$\mathcal{B}(B^+_u\to D^{*-}_sK^+\tau^+\nu_\tau)$                       &$\leq0.31$          &$\cdots $                                                       &$\cdots $                           &$\cdots $               \\\hline
$\mathcal{B}(B^+_u\to \overline{D}^{*0}\pi^0\tau^+\nu_\tau)$            &$\leq0.86$          &$^{2.97\pm0.46_{[D'^{0}_1]}}_{2.59\pm0.38_{[D^{0}_1]}}$         &$2.83\pm0.81_{[D^{*0}_2]}$          &$8.79\pm1.90 $               \\\hline
$\mathcal{B}(B^+_u\to \overline{D}^{*0}\eta\tau^+\nu_\tau)$             &$\leq0.13$          &$\cdots $                                                       &$\cdots $                           &$\cdots $               \\\hline
$\mathcal{B}(B^+_u\to \overline{D}^{*0}\eta'\tau^+\nu_\tau)$            &$\leq0.0088$        &$\cdots $                                                       &$\cdots $                           &$\cdots $               \\\hline
$\mathcal{B}(B^0_d\to \overline{D}^{*0}\pi^-\tau^+\nu_\tau)$            &$\leq1.58$          &$^{5.49\pm0.83_{[D'^{-}_1]}}_{4.79\pm0.74_{[D^{-}_1]}}$         &$5.22\pm1.51_{[D^{*-}_2]}$          &$16.27\pm3.49 $               \\\hline
$\mathcal{B}(B^0_d\to D^{*-}_sK^0\tau^+\nu_\tau)$                       &$\leq0.29$          &$\cdots $                                                       &$\cdots $                           &$\cdots $               \\\hline
$\mathcal{B}(B^0_d\to D^{*-}\pi^0\tau^+\nu_\tau)$                       &$\leq0.79$          &$^{2.73\pm0.41_{[D'^{-}_1]}}_{2.38\pm0.37_{[D^{-}_1]}}$         &$2.56\pm0.74_{[D^{*-}_2]}$          &$8.10\pm1.81 $               \\\hline
$\mathcal{B}(B^0_d\to D^{*-}\eta\tau^+\nu_\tau)$                        &$\leq0.12$          &$\cdots $                                                       &$\cdots $                           &$\cdots$               \\\hline
$\mathcal{B}(B^0_d\to D^{*-}\eta'\tau^+\nu_\tau)$                       &$\leq0.0080$        &$\cdots $                                                       &$\cdots $                           &$\cdots $               \\\hline
$\mathcal{B}(B^0_s\to \overline{D}^{*0}K^-\tau^+\nu_\tau)$              &$\leq0.66$          &$6.10\pm2.64_{[D^{-}_{s1}]}$                                    &$1.06\pm0.37_{[D^{*-}_{s2}]}$       &$7.44\pm3.26 $              \\\hline
$\mathcal{B}(B^0_s\to D^{*-}\overline{K}^0\tau^+\nu_\tau)$              &$\leq0.64$          &$5.08\pm2.33_{[D^{-}_{s1}]}$                                    &$0.80\pm0.28_{[D^{*-}_{s2}]}$       &$6.18\pm2.86 $              \\\hline
$\mathcal{B}(B^0_s\to D_s^{*-}\pi^0\tau^+\nu_\tau)$                     &$\cdots $           &$4.55\pm2.52_{[D'^{-}_{s1}]}$                                   &$\cdots $                           &$\cdots $               \\\hline
$\mathcal{B}(B^0_s\to D_s^{*-}\eta\tau^+\nu_\tau)$                      &$\leq0.17$          &$\cdots $                                                       &$\cdots $                           &$\cdots $                 \\\hline
$\mathcal{B}(B^0_s\to D_s^{*-}\eta'\tau^+\nu_\tau)$                     &$\leq0.019$         &$\cdots $                                                       &$\cdots $                           &$\cdots $               \\\hline
\end{tabular}\label{Tab:BrB2DVPlv}}
\end{table}
Our upper limit predictions of the non-resonant branching ratios are listed in the second column of Tab. \ref{Tab:BrB2DVPlv}. For convenient comparison,  the available experimental values are also given in Tab. \ref{Tab:BrB2DVPlv}.
And we find that only the upper limit of $\mathcal{B}(B^+\to D^{*-}_sK^+\ell'^+\nu_{\ell'})_N$ gives the final effective bound. The upper limit predictions of $\mathcal{B}(B^+_u\to D^{*-}\pi^+\ell'^+\nu_{\ell'})_N$ and  $\mathcal{B}(B^0_d\to \overline{D}^{*0}\pi^-\ell'^+\nu_{\ell'})_N$ are obviously smaller than their experimental upper limits.
Decays $B_{(s)}\to D^*_{(s)}\eta \ell^+\nu_\ell$ and $B_{(s)}\to D^*_{(s)}\eta' \ell^+\nu_\ell$ are suppressed by the narrow phase spaces and the mixing
angle $\theta_P$, and   their upper limits of the branching ratios, especially of $B_{(s)}\to D^*\eta'\tau^+\nu_\tau$,  are obviously smaller than other predictions.


\section{Decays $B \rightarrow  D^* P\ell^+\nu_\ell$  with the $D^{**}$ resonant  states}

Except for the non-resonant $B\to D^*P \ell^+\nu_\ell$ decays, the $B \rightarrow  D^* P\ell^+\nu_\ell$ decay with the $D^{**}$ resonant  state is also studied, where $D^{**}$ is an lightest excited charmed  meson.
There are the four lightest excited charmed meson states, $D^{**}=(D_0, D'_1,D_1,D^*_2)$ or $(D_{s0}, D'_{s1},D_{s1},D^*_{s2})$. The spin-0 states $D_0$ and $D_{s0}$ can only decay to $DP$ \cite{Belle:2022yzd},
so we will  study  $D'_1$, $D_1$ and $D^*_2$ decay to $D^*P$ in this work.

In the case of the decay widths of the resonances are very narrow, the resonant decay branching ratios respect a simple factorization relation
\begin{equation}
\mathcal{B}(B \rightarrow D^{**} \ell^+\nu_\ell, D^{**} \rightarrow D^* P) = \mathcal{B}(B \rightarrow D^{**} \ell^+\nu_\ell) \times  \mathcal{B}(D^{**} \rightarrow D^* P),\label{Eq:Br4BD}
\end{equation}
and this result is also a good approximation for wider resonances. Above Eq. (\ref{Eq:Br4BD}) will be used in our analysis for resonant $B \rightarrow D^{**}(\rightarrow D^* P) \ell^+\nu_\ell$ decays.
Relevant $\mathcal{B}(B \rightarrow D^{**} \ell^+\nu_\ell)$ and $\mathcal{B}(D^{**} \rightarrow D^* P)$  also can be obtained by the SU(3) flavor symmetry.

\subsection{Resonant $B\to D^* P\ell^+\nu_\ell$ decays with the charmed axial vectors }\label{Sec:B2Alv}
\subsubsection{ $B\to D_1/D'_1\ell^+\nu_\ell$ decays }
Similar to $B\to V\ell^+\nu_\ell$ decays, the decay amplitudes of the $B\to A\ell^+\nu_\ell$  decays with $A=D_1/D'_1$ can be written as
\begin{eqnarray}
\mathcal{M}(B\rightarrow A\ell^+\nu_\ell)&=&\frac{G_F}{\sqrt{2}}V_{cb}\sum_{mm'}g_{mm'} L^{\lambda_\ell\lambda_\nu}_mH^{\lambda_A}_{m'},
\end{eqnarray}
with
\begin{eqnarray}
L^{\lambda_\ell\lambda_\nu}_m&=&\epsilon_{\alpha}(m)\bar{\nu_\ell}\gamma^{\alpha}(1-\gamma_5)\ell,\\
H^{\lambda_A}_{m'}&=&\epsilon^*_{\beta}(m')\langle{A}(p,\varepsilon^*)|\bar{c}\gamma^\beta(1-\gamma_5)b|B(p_B)\rangle,
\end{eqnarray}
where the particle helicities $\lambda_A=0,\pm1$, $\lambda_\ell=\pm\frac{1}{2}$ and $\lambda_\nu=+\frac{1}{2}$, as well as $\epsilon_{\mu}(m)$  is the polarization vectors of the virtual $W$ with $m=0,t,\pm1$,
and $\varepsilon^*$ is the polarization vectors of $A$ meson.
Hadronic matrix element can be parameterized by the $B\to A$ form factors  \cite{Cheng:2017pcq,Cheng:2003sm}
\begin{eqnarray}
\left<A(p,\varepsilon^*)\left|\bar{c}\gamma_{\mu}(1-\gamma_5)b\right|B(p_B)\right>
&=&\frac{2iA(q'^2)}{m_B-m_A}\epsilon_{\mu\nu\alpha\beta}\varepsilon^{*\nu}p^\alpha_Bp^\beta\nonumber\\
&&-\left[\varepsilon^*_\mu(m_B-m_A)V^A_1(q'^2)-(p_B+p)_\mu(\varepsilon^*.p_B)\frac{V^A_2(q'^2)}{m_B-m_A}\right]\nonumber\\
&&+q'_\mu(\varepsilon^*.p_B)\frac{2m_A}{q'^2}[V^A_3(q'^2)-V^A_0(q'^2)],
\end{eqnarray}
with $q'=p_{B}-p$.  Then the  hadronic helicity
amplitudes can be written as  \cite{Sun:2011ssd,Li:2009tx,Colangelo:2019axi,Verma:2011yw}
\begin{eqnarray}
H^A_{\pm} &=&(m_{B}-m_A)V^A_1(q'^2)\mp\frac{2m_{B}|\vec{p}_c|}{(m_{B}-m_A)}A(q'^2), \label{Eq:HApm}\\
H^A_{0}&=& \frac{1}{2m_A\sqrt{q'^2}}\left[(m_{B}^2-m_A^2-q'^2)(m_{B}-m_A)V^A_1(q'^2)-\frac{4m_{B}^2|\vec{p}_c|^2}{m_{B}-m_A}V^A_2(q'^2)\right],\label{Eq:HA0} \\
H^A_{t}&=& \frac{2m_{B}|\vec{p}_c|}{\sqrt{q'^2}}V^A_0(q'^2),\label{Eq:HAt}
\end{eqnarray}
where $|\vec{p}_c|\equiv\frac{\sqrt{\lambda}}{2m_{B}}$ with $\lambda=m_B^4+m_A^4+q'^4-2m_B^2m_A^2-2q'^2m^2_B-2q'^2m^2_A$.

The differential branching ratios are  \cite{Ivanov:2019nqd}
\begin{equation}
 \frac{d\mathcal{B}(B\to
  A\ell^+\nu_\ell)}{dq'^2}=\frac{\tau_{B}G_F^2 |V_{cb}|^2\lambda^{1/2}q'^2}{24(2\pi)^3m_{B}^3}\left(1-\frac{m_\ell^2}{q'^2}\right)^2\left(1+\frac{m_\ell^2}{2q'^2}\right)
 {\cal H}_{\rm total}(B\to A\ell^+\nu_\ell),\label{Eq:dbdq2}
\end{equation}
with
\begin{eqnarray}
{\cal H}_{\rm total}(B\to A\ell^+\nu_\ell)&=& \Big(\big|H^A_+\big|^2+\big|H^A_-\big|^2+\big|H^A_0\big|^2\Big) +\frac{3m_\ell^2}{2q'^2}\Big/\left(1+\frac{m_\ell^2}{2q'^2}\right)\big|H^A_t\big|^2,  \label{eq:hh}
  \end{eqnarray}
where  $m_\ell^2\leq q'^2\leq(m_{B}-m_A)^2$.
Then one can obtain the branching ratios by the Eq. (\ref{Eq:dbdq2}) and the form factors. Usually, it is difficult to calculate the form factors, and their results depend on the different approaches.

The  branching ratios also can be obtained by using the SU(3) flavor symmetry.
In the $\ell=e,\mu$ cases,  $\frac{3m_\ell^2}{2q'^2}$ is far less than  $1+\frac{m_\ell^2}{2q'^2}$  in Eq. (\ref{eq:hh}), so the $\big|H^A_t\big|^2$ term in Eq.(\ref{eq:hh}) can ignore safely.   In the $\ell=\tau$ case,
$(\frac{3m_\ell^2}{2q'^2})/(1+\frac{m_\ell^2}{2q'^2})$ lie in $[0.48,1]$, the $\big|H^A_t\big|^2$ term give non-negligible contribution, but it is difficult to estimate its contribution
if we do not depend on any calculation of the form  factors.  Two cases will be considered
in our analysis of the $B\to D_1/D'_1\ell^+\nu_\ell$ decays.
\begin{itemize}
\item[{\bf $S_1$}:] Ignoring $\big|H^A_t\big|^2$ term in the $\ell=e,\mu,\tau$ decays,  then ${\cal H}_{\rm total}$ only includes the hadronic part that  can be  related by the SU(3) flavor symmetry as follow.
\begin{eqnarray}
{\cal H}_{\rm total}(B^+_u\to\overline{D}'^0_1\ell^+\nu_{\ell})&=&{\cal H}_{\rm total}(B^0_d\to D'^-_1\ell'^+\nu_{\ell'})={\cal H}_{\rm total}(B^0_s\to D'^-_{s1}\ell'^+\nu_{\ell'})=|a_0|^2,\label{Eq:Re1}\\
{\cal H}_{\rm total}(B^+_u\to\overline{D}^0_1\ell^+\nu_{\ell})&=&{\cal H}_{\rm total}(B^0_d\to D^-_1\ell^+\nu_{\ell})={\cal H}_{\rm total}(B^0_s\to D^-_{s1}\ell^+\nu_{\ell})=|b_0|^2.\label{Eq:Re2}
\end{eqnarray}
where $a_0$ and $b_0$ are the nonperturbative coefficients. The actual  $\mathcal{B}(B\to  A\tau^+\nu_\tau)$ might be larger than our later predictions.

\item[{\bf $S_2$}:] ${\cal H}_{\rm total}$  are obtained by using the hadronic helicity amplitude expressions in Eqs. (\ref{Eq:HApm}-\ref{Eq:HAt}), which are $q^2$ dependent and can be expressed by the
 form factors. The form factors of $B\to D^{1/2}_1,D^{3/2}_1$ in Ref. \cite{Verma:2011yw} are taken (we do not use ones of $B_s\to D^{1/2}_{s1},D^{3/2}_{s1}$), we keep $V^A_1(0)$ as an undetermined constant,
 other $F_i(0)$ can be expressed as  $r_F\times V^A_1(0)$, and  $r_F=\frac{F_i(0)}{V^A_1(0)}$ are taken from Ref. \cite{Verma:2011yw}.
 Since these form factors also preserve the SU(3) flavor symmetry, the same relations in Eqs. (\ref{Eq:Re1}-\ref{Eq:Re2}) will be used for $V^A_1(0)$.

\end{itemize}

For the semileptonic $B\to D^{(')}_1\ell^+\nu_\ell$ decays, the branching ratios of $B^+_u\to\overline{D}'^0_1\ell'^+\nu_{\ell'}$, $B^0_d\to D'^-_1\ell'^+\nu_{\ell'}$, $B^+_u\to\overline{D}^0_1\ell'^+\nu_{\ell'}$, and $B^0_d\to D^-_1\ell'^+\nu_{\ell'}$ decays
have been measured by the Belle Collaboration \cite{Belle-II:2022evt},  and the experimental data with $2\sigma$ errors are listed in the second column of Tab. \ref{Tab:BrDAB2Plv}.
Using the experimental data of $\mathcal{B}(B^+_u\to\overline{D}'^0_1\ell'^+\nu_{\ell'})$ and $\mathcal{B}(B^0_d\to D'^-_1\ell'^+\nu_{\ell'})$,  one can constrain the allowed range of $a_0$ and $V^{D'_1}_1(0)$,
and then one can obtain the predictions of $\mathcal{B}(B^+_u\to\overline{D}'^0_1\tau^+\nu_{\tau})$,
$\mathcal{B}(B^0_d\to D'^-_1\tau^+\nu_{\tau})$ and $\mathcal{B}(B^0_s\to D'^-_{s1}\ell^+\nu_{\ell})$ by the constrained $a_0$ or $V^{D'_1}_1(0)$. In a similar way, one can constrain $b_0$ and $V^{D_1}_1(0)$, and predict $\mathcal{B}(B^+_u\to\overline{D}^0_1\tau^+\nu_{\tau})$,
$\mathcal{B}(B^0_d\to D^-_1\tau^+\nu_{\tau})$ and $\mathcal{B}(B^0_s\to D^-_{s1}\ell^+\nu_{\ell})$.
 We obtain that $|a_0|=4.50\pm0.52$ and $|b_0|=5.08\pm1.05$ in the $S_1$ case, and  $|V^{D'_1}_1(0)|=0.57\pm0.32$ and $|V^{D_1}_1(0)|=0.51\pm0.15$ in the $S_2$ case.
In Ref. \cite{Verma:2011yw},  $|V^{D'_1}_1(0)|=0.19\pm0.02\pm0.01$ and $|V^{D_1}_1(0)|=0.58\pm0.01^{+0.02}_{-0.03}$.  Our constrained  $|V^{D'_1}_1(0)|$ is  larger than one in Ref. \cite{Verma:2011yw}, but it  close to the latter within $2\sigma$ errors.  Our constrained  $|V^{D_1}_1(0)|$ is  consistent with  one in Ref. \cite{Verma:2011yw}.

Note that, if considering the SU(3) flavor breaking effects from different masses of $u$, $d$, and $s$ quarks, the  nonperturbative coefficients of the $B^0_s\to D^{(')-}_{s1}\ell^+\nu_{\ell}$ decays are different from ones of the  $B^+_u\to\overline{D}^{(')0}_1\ell^+\nu_{\ell}$ and $B^0_d\to D^{(')-}_1\ell^+\nu_{\ell}$. After the $B^0_s\to D^{(')-}_{s1}\ell^+\nu_{\ell}$ decays are measured, one can estimate the  SU(3) flavor breaking effects in the $B\to D^{(')}_1\ell^+\nu_\ell$ decays. The same  situation also appears in the later $B\to D^*_2\ell^+\nu_\ell$ decays.

Our predictions in the  $S_1$ and $S_2$ cases are listed in Tab. \ref{Tab:BrDAB2Plv}.
Comparing the $\mathcal{B}(B\to D^{(')}_1\ell'^+\nu_{\ell'})$ predictions in  $S_1$ to ones in $S_2$, $\mathcal{B}(B^+_u\to\overline{D}'^0_1\ell'^+\nu_{\ell'})$, $\mathcal{B}(B^0_d\to D'^-_1\ell'^+\nu_{\ell'})$, $\mathcal{B}(B^+_u\to\overline{D}^0_1\ell'^+\nu_{\ell'})$
 and $\mathcal{B}(B^0_d\to D^-_1\ell'^+\nu_{\ell'}$
are exactly same in both $S_1$ and $S_2$.   $\mathcal{B}(B^0_s\to D'^-_{s1}\ell'^+\nu_{\ell'})$ and $\mathcal{B}(B^0_s\to D^-_{s1}\ell'^+\nu_{\ell'})$  are slightly difference in two cases due to different $q'^2$ dependence.
Comparing our SU(3) flavor symmetry predictions of $\mathcal{B}(B\to D^{(')}_1\ell'^+\nu_{\ell'})$ with their  experimental data, one can see that our predictions are coincident with the present data.
As given in the sixth and seventh columns of Tab. \ref{Tab:BrDAB2Plv},  the central values of $\mathcal{B}(B\to D^{(')}_1\tau^+\nu_{\tau})$ predictions in  $S_1$ to ones in $S_2$ are similar to each other,
nevertheless, the errors of the predictions, specially for $\mathcal{B}(B\to D'_1\tau^+\nu_{\tau})$, in $S_2$ case are obviously larger than ones in $S_1$ case.
In later analysis, we will use the predictions of $\mathcal{B}(B\to D^{(')}_1\ell^+\nu_\ell)$ in $S_1$ case to give the numerical results of  $\mathcal{B}(B\to D^* P\ell^+\nu_\ell)$  with the $D'_1/D_1$ resonant states.

\begin{table}[t]
\renewcommand\arraystretch{1.5}
\tabcolsep 0.05in
\centering
\caption{ Experimental data and predictions of the branching ratios of  the semileptonic  $B\to D^{(')}_1\ell^+\nu_\ell$ decays with the $2\sigma$ errors by  the SU(3) flavor symmetry(in units of $10^{-3}$).
}\vspace{0.1cm}
{\footnotesize
\begin{tabular}{lccc|lcc}  \hline\hline
                                                                 & Data \cite{Belle-II:2022evt}    & Predictions in $S_1$  & Predictions in $S_2$  &                                                                & Predictions in $S_1$       & Predictions in $S_2$           \\\hline
$\mathcal{B}(B^+_u\to\overline{D}'^0_1\ell'^+\nu_{\ell'})$       &$4.2\pm1.8$                           &$5.18\pm0.70$          &$5.18\pm0.70$          &  $\mathcal{B}(B^+_u\to\overline{D}'^0_1\tau^+\nu_\tau)$          &$0.88\pm0.13$             &$0.87\pm0.65$                  \\
$\mathcal{B}(B^0_d\to D'^-_1\ell'^+\nu_{\ell'})$                &$3.9\pm1.6$                            &$4.85\pm0.65$          &$4.85\pm0.65$          &  $\mathcal{B}(B^0_d\to D'^-_1\tau^+\nu_\tau)$                    &$0.82\pm0.12$             &$0.81\pm0.61$                  \\
$\mathcal{B}(B^0_s\to D'^-_{s1}\ell'^+\nu_{\ell'})$             &$\cdots$                               &$4.99\pm0.79$          &$5.15\pm0.82$          &  $\mathcal{B}(B^0_s\to D'^-_{s1}\tau^+\nu_\tau)$                 &$0.89\pm0.14$             &$0.90\pm0.67$                  \\
%
$\mathcal{B}(B^+_u\to\overline{D}^0_1\ell'^+\nu_{\ell'})$       &$6.6\pm2.2$                            &$6.65\pm2.15$          &$6.65\pm2.15$          &  $\mathcal{B}(B^+_u\to\overline{D}^0_1\tau^+\nu_\tau)$           &$1.13\pm0.38$             &$1.04\pm0.55$                 \\
$\mathcal{B}(B^0_d\to D^-_1\ell'^+\nu_{\ell'})$                 &$6.2\pm2.0$                            &$6.20\pm2.00$          &$6.20\pm2.00$          &  $\mathcal{B}(B^0_d\to D^-_1\tau^+\nu_\tau)$                     &$1.05\pm0.35$             &$0.97\pm0.51$                  \\
$\mathcal{B}(B^0_s\to D^-_{s1}\ell'^+\nu_{\ell'})$              &$\cdots$                               &$6.29\pm2.11$          &$6.60\pm2.23$          &  $\mathcal{B}(B^0_s\to D^-_{s1}\tau^+\nu_\tau)$                  &$1.08\pm0.36$             &$1.06\pm0.56$                  \\\hline
\end{tabular}\label{Tab:BrDAB2Plv}}
\end{table}

\subsubsection{ $D_1/D'_1\to  D^*P$ decays }

In the SU(3) flavor symmetry limit, the decay amplitudes of the  $D^{(')}_1\to D^*P$ decays, which decay through the strong  or  electromagnetic interactions,   can be simply parametrized as
\begin{eqnarray}
A(D'_1\to D^*P) = a^A D'^i_1D^*_jP^j_i,~~~~~~~~~A(D_1\to D^*P) = a^B D_1^iD^*_jP^j_i,\label{Eq:AD12DVP}
\end{eqnarray}
where $a^{A,B}$ are the nonperturbative coefficients. The amplitude relations of the  $D^{(')}_1\to D^*P$ decays are given  in Tab. \ref{Tab:DA2DVPAmp}.

For the $D'_1\to D^* P$ decays, they have not been observed yet. Decays $D'_1\to D^*\eta$ and $D'_1\to D^*\eta'$ are not allowed by the phase spaces.
So  we assume $\mathcal{B}(D'_1\to D^* \pi)=1$  and use the amplitude relations given in Tab.  \ref{Tab:DA2DVPAmp} to obtain each branching ratios of the $D'_1\to D^* \pi$ decays,
which are listed in the second column of Tab. \ref{Tab:BrD12DVP}.
Decays $D'^{+}_{s1}\to D^{*0}K^+$ and $D'^{+}_{s1}\to D^{*+}K^0$  are also not allowed by the phase spaces. Decay  $D'^+_{s1} \to D^{*+}_s\pi^0$ is decayed by different way, its amplitude can not be related by nonperturbative coefficient $a^A$.
The experimental data $\mathcal{B}(D'^+_{s1}\to D^{*+}_s\pi^0)=(48\pm22)\%$ from PDG \cite{ParticleDataGroup:2022pth} will be used in later  prediction of $\mathcal{B}(B^0_s\to D'^-_{s1}\ell^+\nu_\ell,D'^-_{s1}\to D^{*+}_s\pi^0)$.

\begin{table}[t]
\renewcommand\arraystretch{1.5}
\tabcolsep 0.3in
\centering
\caption{Decay amplitudes for the $D^{(')}_1\to D^*P$ decays  by  the SU(3) flavor symmetry.
}\vspace{0.0cm}
{\footnotesize
\begin{tabular}{lc|lc}  \hline\hline
Decay amplitudes  & SU(3)  amplitudes                 &Decay amplitudes                & SU(3) amplitudes                                                                                                 \\\hline
$A(D'^{0}_1\to D^{*0}\pi^0)$&$a^A/\sqrt{2}$           &$A(D^{0}_1\to D^{*0}\pi^0)$     &$a^B/\sqrt{2}$            \\
$A(D'^{0}_1\to D^{*+}\pi^-)$&$a^A $                   &$A(D^{0}_1\to D^{*+}\pi^-)$     &$a^B $                    \\
$A(D'^{+}_1\to D^{*+}\pi^0)$&$-a^A/\sqrt{2}$          &$A(D^{+}_1\to D^{*+}\pi^0)$     &$-a^B/\sqrt{2}$            \\
$A(D'^{+}_1\to D^{*0}\pi^+)$&$a^A $                   &$A(D^{+}_1\to D^{*0}\pi^+)$     &$a^B $                       \\           %
$A(D'^{+}_{s1}\to D^{*0}K^+)$&$a^A$                   &$A(D^{+}_{s1}\to D^{*0}K^+)$    &$a^B$                         \\
$A(D'^{+}_{s1}\to D^{*+}K^0)$&$a^A $                  &$A(D^{+}_{s1}\to D^{*+}K^0)$    &$a^B $                     \\ \hline
\end{tabular}\label{Tab:DA2DVPAmp}}
\renewcommand\arraystretch{1.4}
\tabcolsep 0.3in
\centering
\caption{Branching ratios of the $D^{(')}_1\to D^*P$ decays with the $2\sigma$ errors (in units of $10^{-2}$). }\vspace{0.1cm}
{\footnotesize
\begin{tabular}{lc|lc}  \hline\hline
Decay modes                        & Predictions           &Decay modes                      & Predictions        \\\hline
 $D'^{0}_1\to D^{*0}\pi^0$         &$33.65 \pm0.03 $       &$D^{0}_1\to D^{*0}\pi^0$         &$26.71 \pm11.44 $           \\
 $D'^{0}_1\to D^{*+}\pi^-$         &$66.35 \pm0.03 $       &$D^{0}_1\to D^{*+}\pi^-$         &$52.73 \pm22.61 $           \\
  $D'^{+}_1\to D^{*+}\pi^0$         &$33.23\pm0.01 $       &$D^{+}_1\to D^{*+}\pi^0$         &$26.48 \pm11.35 $             \\
  $D'^{+}_1\to D^{*0}\pi^+$         &$66.77 \pm0.01 $     &$D^{+}_1\to D^{*0}\pi^+$          &$53.19 \pm22.79 $              \\
  $$                                &$$                    &$D^{+}_{s1}\to D^{*0}K^+$        &$54.98 \pm7.13$                \\
  $$                                &$$                    &$D^{+}_{s1}\to D^{*+}K^0$        &$45.02\pm7.13$                  \\\hline
\end{tabular}\label{Tab:BrD12DVP}}
\end{table}

$D_1$ decays are dominant by $D_1\to D^*\pi$ and $D_1\to D\pi\pi$.
We use $\mathcal{B}(D_1^-\to \overline{D}^{*0}\pi^-)=\frac{\mathcal{B}(B^0\to D_1^-\ell\nu_\ell, D_1^-\to \overline{D}^{*0}\pi^-)}{\mathcal{B}(B^0\to D_1^-\ell\nu_\ell)}$ to determine the nonperturbative coefficient $a^B$,
and then give the branching ratios of every relevant decay mode. The results of $D^0_1$ and $D^+_1$ decays are listed in the forth  column of Tab. \ref{Tab:BrD12DVP}.
The decay width of $D_{s1}^+$, $\Gamma_{D_{s1}^+}=(0.92\pm0.10)\times10^{-3}$, is much smaller than ones of $D_{1}^{0,+}$,     $\Gamma_{D_{1}^{0,+}}=(31.3\pm3.8)\times10^{-3}$.
The predictions of $\mathcal{B}(D^{+}_{s1}\to D^{*0}K^+)$ and $\mathcal{B}(D^{+}_{s1}\to D^{*+}K^0)$  are much larger than $100\%$ by the constrained $a^B$ from $\mathcal{B}(D_1^-\to \overline{D}^{*0}\pi^-)$,
they are  not right,  and we will not use them.
Using $\frac{\mathcal{B}(D_{s1}^+\to \overline{D}^{*+}_sK^0)}{\mathcal{B}(D_{s1}^+\to \overline{D}^{*+}K^0)}=0.85\pm0.24$ from PDG \cite{ParticleDataGroup:2022pth} and
assuming  $\mathcal{B}(D_{s1}^+\to \overline{D}^{*+}_sK^0)+\mathcal{B}(D_{s1}^+\to \overline{D}^{*+}K^0)=1$,   these branching ratios are obtained, which are given in the forth  column of Tab. \ref{Tab:BrD12DVP}.

\subsubsection{ $\mathcal{B}(B\to D^* P\ell^+\nu_\ell)$  with the $D'_1/D_1$ resonant states }
In terms of $\mathcal{B}(B\to D^{(')}_1\ell^+\nu_\ell)$ in the $S_1$ case given in Tab. \ref{Tab:BrDAB2Plv}   and  $\mathcal{B}(D^{(')}_1\to D^*P)$ given in Tab. \ref{Tab:BrD12DVP},  after considering relevant  experimental bounds in Eqs. (\ref{EBrDABu2DVpilv}-\ref{EBrDApBu2DVpilv})
 and Eqs. (\ref{EBrDABd2DVpilv}-\ref{EBrDApBd2DVpilv}),
one can obtain $\mathcal{B}(B\to D^* P\ell^+\nu_\ell )$ with the resonant charmed axial vector. The results are summarized in the third  column of Tab. \ref{Tab:BrB2DVPlv}. There are four experimental data  for $\mathcal{B}(B^+_u\to D^{*-}\pi^+\ell'^+\nu_{\ell'})_{[D^{'0}_1,D^0_1]}$ and  $\mathcal{B}(B^0_d\to \overline{D}^{*0}\pi^-\ell'^+\nu_{\ell'})_{[D^{'-}_1,D^-_1]}$, which are also listed in the third  column of Tab. \ref{Tab:BrB2DVPlv}.    Comparing with the data, one can see that $\mathcal{B}(B^+_u\to D^{*-}\pi^+\ell'^+\nu_{\ell'})_{[D^{'0}_1,D^0_1]}$  give the further effective constraints. The predictions of  $\mathcal{B}(B^0_d\to \overline{D}^{*0}\pi^-\ell'^+\nu_{\ell'})_{[D^{'-}_1,D^-_1]}$ are consistent with their experimental data but with smaller errors.


\subsection{Resonant $B\to D^* P\ell^+\nu_\ell$ decays with the charmed  tensors }

For the semileptonic $B\to D^*_2\ell^+\nu_\ell$ decays, the branching ratios of $B^+_u\to\overline{D}^{*0}_2\ell'^+\nu_{\ell'}$ and  $B^0_d\to D^{*-}_2\ell'^+\nu_{\ell'}$ decays
have been measured by the Belle Collaboration \cite{Belle-II:2022evt},  and the experimental data with $2\sigma$ errors are listed in the second column of Tab. \ref{Tab:BrB2DTlv}.
In similar to $S_2$ case in Sec. \ref{Sec:B2Alv}, the SU(3) flavor symmetry predictions of the non-measured $B\to D^*_2\ell^+\nu_\ell$  decays
from the two experimental data are obtained in Ref. \cite{WangB2DPlv2023}, and they are listed in the third column of Tab. \ref{Tab:BrB2DTlv}.

For the decay amplitudes of the $D^*_2\to D^*P$  decays, they are similar to  ones of the  $D_1\to D^*P$ decays in Eq. (\ref{Eq:AD12DVP}) in terms of  replacing $a^{B}$ by $a^T$  and replacing  $D_1$ by $D^*_2$.
The decay amplitudes of the specific $D^*_2\to D^*P$ decays are similar to ones in Tab. \ref{Tab:DA2DVPAmp}, so we will not show them here.
Decays $D^*_2\to D^*P$ and $D^*_2\to DP$  have not been measured up to now, therefore, we can not constrain $a^T$ directly.  In Ref. \cite{Belle-II:2022evt}, they make a assumption that $\mathcal{B}(D^{*}_2\to D^*\pi)+\mathcal{B}(D^{*}_2\to D\pi)=1$.
Using $\frac{\mathcal{B}(D^*_2\to D\pi)}{\mathcal{B}(D^*_2\to D^*\pi)}=1.52\pm0.14$  within $2\sigma$  errors from PDG \cite{ParticleDataGroup:2022pth}, and  assuming  one of $\mathcal{B}(D^{*0}_2\to D\pi,D^*\pi)$,
$\mathcal{B}(D^{*+}_2\to D\pi,D^*\pi)$ and $\mathcal{B}(D^{*+}_{s2}\to DK,D^*K)$ is equal to one and other two values are less than or equal to one,  one  can  constrain  on the nonperturbative coefficients $a^T$ and $b^T$ (where $b^T$ is similar to $a^T$ but for the  $D^*_2\to DP$ decays), and they are  $|a^T|=9.06\pm0.85$ and $|b^T|=25.14\pm1.47$.
 And then, the branching ratios of the $D^*_2\to D^*P$  decays can be predicted,   which are summarized in the second column of Tab. \ref{Tab:BrT2DVP}.
In addition,  their decay width predictions  and previous width predictions are also given in the forth and fifth  columns of Tab. \ref{Tab:BrT2DVP}, respectively.  Our width predictions are larger than ones in Ref. \cite{Li:2015xka}.
\begin{table}[t]
\renewcommand\arraystretch{1.42}
\tabcolsep 0.25in
\centering
\caption{Branching ratios of the  $B\to D^*_2\ell^+\nu_\ell$ decays with the $2\sigma$ errors.
}\vspace{0.1cm}
{\footnotesize
\begin{tabular}{lcc}  \hline\hline
                                            & Exp. data \cite{Belle-II:2022evt}     & Predictions     \cite{WangB2DPlv2023}                      \\\hline
$\mathcal{B}(B^+_u\to\overline{D}^{*0}_2\ell'^+\nu_{\ell'})(\times10^{-3})$        &$2.9\pm0.6$              &$3.20\pm0.30$\\
$\mathcal{B}(B^0_d\to D^{*-}_2\ell'^+\nu_{\ell'})(\times10^{-3})$                  &$2.7\pm0.6$              &$2.98\pm0.29$\\
$\mathcal{B}(B^0_s\to D^{*-}_{s2}\ell'^+\nu_{\ell'})(\times10^{-3})$               &$\cdots$                 &$2.72\pm0.27$\\
$\mathcal{B}(B^+_u\to\overline{D}^{*0}_2\tau^+\nu_\tau)(\times10^{-4})$            &$\cdots$                 &$2.15\pm0.47$\\
$\mathcal{B}(B^0_d\to D^{*-}_2\tau^+\nu_\tau)(\times10^{-4})$                      &$\cdots$                 &$1.97\pm0.43$\\
$\mathcal{B}(B^0_s\to D^{*-}_{s2}\tau^+\nu_\tau)(\times10^{-4})$                   &$\cdots$                  &$1.73\pm0.38$ \\\hline
\end{tabular}\label{Tab:BrB2DTlv}}
\renewcommand\arraystretch{1.42}
\tabcolsep 0.1in
\centering
\caption{Branching ratios of the  $T\to D^*P$ decays with the $2\sigma$ errors.
}\vspace{0.1cm}
{\footnotesize
\begin{tabular}{lccc}  \hline\hline
Decay modes  & Branching ratios $(\times10^{-2})$ & Decay widths (MeV) &  Decay widths from Ref. \cite{Li:2015xka}  (MeV)   \\\hline
$D^{*0}_2(2460)\to D^{*0}\pi^0$         &$12.96 \pm2.29 $                   &$6.19 \pm1.12 $                   & $2.40^{+1.74}_{-0.94}$\\
$D^{*0}_2(2460)\to D^{*+}\pi^-$         &$24.56 \pm4.36 $                   &$11.71 \pm2.13 $                   & $3.99^{+1.22}_{-1.56} $\\
$D^{*+}_2(2460)\to D^{*+}\pi^0$         &$12.49 \pm2.09 $                   &$5.98 \pm1.04 $                   & $\cdots$\\
$D^{*+}_2(2460)\to D^{*0}\pi^+$         &$25.49 \pm4.25 $                   &$12.21 \pm2.13 $                   & $\cdots$\\
$D^{*+}_{s2}(2573)\to D^{*0}K^+$           &$6.07 \pm1.54 $                   &$1.02 \pm0.22 $                   & $0.20^{+0.09}_{-0.07}$\\
$D^{*+}_{s2}(2573)\to D^{*+}K^0$           &$4.60 \pm1.19 $                   &$0.78 \pm0.17 $                   & $0.15^{+0.06}_{-0.05}$\\\hline
\end{tabular}\label{Tab:BrT2DVP}}
\end{table}

Using  $\mathcal{B}(B\to D^*_2\ell^+\nu_\ell)$ given in Tab. \ref{Tab:BrB2DTlv}   and  $\mathcal{B}(D^*_2\to D^*P)$ given in Tab. \ref{Tab:BrT2DVP},
after considering relevant  experimental bounds in Eq. (\ref{EBrDTBu2DVpilv}) and Eq. (\ref{EBrDApBu2DVpilv}),
one can obtain $\mathcal{B}(B\to D^* P\ell^+\nu_\ell )$ with the charmed  tensor, and they are summarized in the forth column of Tab. \ref{Tab:BrB2DVPlv}.
 Comparing with the data of $\mathcal{B}(B^+_u\to D^{*-}\pi^+\ell'^+\nu_{\ell'})$  and $\mathcal{B}(B^0_d\to \overline{D}^{*0}\pi^-\ell'^+\nu_{\ell'})$, one can see the upper limit of $\mathcal{B}(B^0_d\to \overline{D}^{*0}\pi^-\ell'^+\nu_{\ell'})$ gives further constraint on the relevant nonperturbative parameters and the branching ratio predictions.

\section{Global analysis}
In this section, we will analyse the numerical results of $\mathcal{B}(B\to D^* P\ell^+\nu_\ell)$  given in Tab. \ref{Tab:BrB2DVPlv}.
As given in Tab. \ref{Tab:BrB2DVPlv}, we can see that $\mathcal{B}(B\to D^* P\tau^+\nu_{\tau})$ with the non-resonant and the charmed axial vector resonant contributions  are much smaller than corresponding $\mathcal{B}(B\to D^* P\ell'^+\nu_{\ell'})$
 since the former small phase spaces.

We can see that $B^{0,+}\to D^*\eta\ell^+\nu_\ell$, $B^{0,+}\to D^*\eta'\ell^+\nu_\ell$,
$B^{0}_s\to D^*_s \eta\ell^+\nu_\ell$, $B^{0}_s\to D^*_s\eta'\ell^+\nu_\ell$ and $B^{0,+}\to D^{*}_sK\ell^+\nu_\ell$  decays
 only receive the non-resonant contributions. Under the SU(3) flavor symmetry, $\mathcal{B}(B^0_s\to D_s^{*-}\pi^0\ell^+\nu_{\ell})_N=0$, and $\mathcal{B}(B^0_s\to D_s^{*-}\pi^0\ell^+\nu_{\ell})_{all}$
 only include the $D'^-_{s1}$ resonant contributions.

Other $B\to D^* P\ell^+\nu_\ell$ decays  receive the contributions of the non-resonant
states, the charmed axial vector resonant  states  and  the charmed  tensor resonant
states. In the $\ell=e,\mu$ decays,  the charmed axial vector resonant contributions are largest. But in  the $\ell=\tau$ decays, both the charmed axial vector and tensor resonant contributions are important.
We sum all three kinds of contributions as $\mathcal{B}_{all}$ (for the  non-resonant branching ratios, they are taken in the ranges of 0 to their maximal values listed in the second column of Tab. \ref{Tab:BrB2DVPlv}.), after considering the experimental bounds of $\mathcal{B}(B^+\to D^{*-}\pi^+\ell'^+\nu_{\ell'})_{all}$  in Eq. (\ref{EBrTBu2DVpilv})
and $\mathcal{B}(B^0\to D^{*0}\pi^+\ell'^+\nu_{\ell'})_{all}$ in Eq. (\ref{EBrTBd2DVpilv}), the results of $\mathcal{B}_{all}$ are   listed in the last column of Tab. \ref{Tab:BrB2DVPlv}.

Noted that the interference terms between non-resonant, the axial vector resonant and  the  tensor resonant contributions  exist, but they have not been considered in this work,
and they might not be ignored if all three kinds of contributions or  two kinds of contributions  are important in the decays. And they will be studied in our succeeding work.

\section{Summary}
In this paper, we have studied the $B\to D^*P\ell^+\nu_\ell$  decay processes with  the non-resonant, the charmed axial vector resonant and  the charmed  tensor resonant contributions
by using the SU(3) flavor symmetry.
The amplitude relations of the non-resonant $B\to D^*P\ell^+\nu_\ell$ decays, the semileptonic $B\to D^{**}\ell^+\nu_\ell$ decays and the non-leptonic $D^{**}\to D^*P$ decays  have been obtained by the SU(3) flavor symmetry.
And then using relevant  experimental data of the $B\to D^*P\ell'^+\nu_{\ell'}$ decays, we have presented a theoretical
analysis of these decays. Our main results can be summarized as follows.

We have found that  $B^{0,+}\to D^*\eta\ell^+\nu_\ell$, $B^{0,+}\to D^*\eta'\ell^+\nu_\ell$, $B^{0}_s\to D^*_s \eta\ell^+\nu_\ell$, $B^{0}_s\to D^*_s\eta'\ell^+\nu_\ell$ and $B^{0,+}\to D^{*}_sK\ell^+\nu_\ell$ decays only receive
non-resonant contributions. Decays $B^0_s\to D_s^{*-}\pi^0\ell^+\nu_\ell$ only  receive the $D'_{s1}$ resonant contributions,  and the latter contributions might be larger than the former.
Other decays receive the non-resonant, the charmed axial vector resonant and  the charmed  tensor resonant contributions, all three kinds of contributions are important in most of decays.
Many branching ratios have been predicted on the order of $\mathcal{O}(10^{-3})$, and they might be measured in future experiments.

Although only  approximate predictions can be obtained by the SU(3) flavor symmetry, they are still useful for understanding these decays.
So far, our predictions under the SU(3) flavor symmetry are  quite   agree with  present experimental data.
Our predictions could be tested in future experiments, such as LHCb  and Belle II.

\section*{ACKNOWLEDGEMENTS}
The work was supported by the National Natural Science Foundation of China (No. 12365014 and No. 12175088).

\end{document}